\documentclass{iaus}
\usepackage{graphicx}

\def\pa{\partial}
\def\vb{{\bf v}}
\def\Bb{{\bf B}}
\def\ol{\overline}

\title[] 
{Flux-transport and mean-field dynamo theories of solar cycles}

\author[Choudhuri]   
{Arnab Rai Choudhuri}

\affiliation{Department of Physics, Indian Institute of Science, Bangalore-560012\\ email: {\tt arnab@physics.iisc.ernet.in}} 

\pubyear{2010}
\volume{xxx}  
\pagerange{1--6}
\jname{Title of your IAU Symposium}
\editors{A.C. Editor, B.D. Editor \& C.E. Editor, eds.}
\begin{document}

\maketitle

\begin{abstract}
We point out the difficulties in carrying out direct numerical simulation
of the solar dynamo problem and argue that kinematic mean-field models
are our best theoretical tools at present for explaining various aspects
of the solar cycle in detail.
The most promising kinematic mean-field model 
is the flux transport dynamo model, in which the toroidal field is
produced by differential rotation in the tachocline, the poloidal
field is produced by the Babcock--Leighton mechanism at the solar
surface and the meridional circulation plays a crucial role.  Depending
on whether the diffusivity is high or low, either the diffusivity or
the meridional circulations provides the main transport mechanism for
the poloidal field to reach the bottom of the convection zone from the
top.  We point out that the high-diffusivity flux transport dynamo model
is consistent with various aspects of observational data. The irregularities
of the solar cycle are primarily produced by fluctuations in the 
Babcock--Leighton mechanism and in the meridional circulation. We
summarize recent work on the fluctuations of meridional circulation
in the flux transport dynamo, leading to explanations of such things
as the Waldmeier effect.

\end{abstract}

\firstsection 
\section{In defence of kinematic mean-field models}

We believe that the solar magnetic fields are generated by the
interactions between the velocity field $\vb$ and the magnetic field $\Bb$ within the
Sun's convections zone. These interactions are described by
the following MHD equations:
$$\frac{\pa \vb}{\pa t} + (\vb.\nabla) \vb = - \frac{1}{\rho}
\nabla \left( p + \frac{B^2}{8 \pi} \right) +  \frac{(\Bb. \nabla)\Bb}
{4 \pi \rho} + {\bf g} + \nu \, \nabla^2 \vb, \eqno(1)$$
$$\frac{\pa \Bb}{\pa t} = \nabla \times (\vb \times \Bb)
+ \lambda \, \nabla^2 \Bb. \eqno(2)$$
Here $\rho$ is density, $p$ is pressure, ${\bf g}$ is gravitational
field, $\nu$ is kinematic viscosity and 
$\lambda$ is magnetic diffusivity.

There are two possible approaches to the solar dynamo problem for
explaining the generation of solar magnetic fields.

\begin{itemize}

\item Direct numerical simulation or DNS, in which one puts all the
MHD equations in the computer and solves them.

\item The kinematic approach, in which the velocity field is given 
and one solves only the equation (2) for the magnetic field.

\end{itemize}

\noindent Historically, solar dynamo theory developed by following
the kinematic approach.  Since the turbulence in the solar convection
zone is crucial for the dynamo action, the kinematic theory has to
be of the nature of a mean field theory in which we average over
fluctuations around the mean (Parker 1955; Steenbeck, Krause \&
R\"adler 1966). We can split 
both the velocity field and the magnetic field into mean and
fluctuating parts, i.e. 
$$\vb = \ol{\vb} + \vb', \; \; \; \Bb =\ol{\Bb} + \Bb'. \eqno(3)$$
Here the overline indicates the mean and the prime indicates the
departure from the mean. On substituting (3) in (2), one finds that
$\ol{\Bb}$ obeys an equation more complicated than (2) obeyed by
$\Bb$ --- the celebrated dynamo equation (see, for example,
Choudhuri 1998; \S~16.5). Kinematic mean-field models
are based on this dynamo equation.

Those of us who work on kinematic mean-field models of the solar
dynamo nowadays often face a question: ``Maybe the kinematic 
mean-field approach was the only possible approach when Parker was
developing the subject. But is it not old-fashioned and out-of-date
to still work on kinematic mean-field models?  Why don't you do DNS?"
The first attempt of doing a DNS of the solar dynamo was made quite
early --- by Gilman (1983). But Gilman's simulations failed to
reproduce even the most basic features of the solar cycle, and the
reasons for this failure are still not fully understood. Perhaps
this early failure discouraged further work on DNS of the solar
dynamo for a long time.

Impressive simulations of the geodynamo are now available --- beginning
with the work of Glatzmaier \& Roberts (1995). Even the random
reversals of the geomagnetic field had been successfully modelled.
Doing a DNS of the solar dynamo is much more challenging for the
following reasons.
\begin{itemize}
\item The solar convection zone is highly stratified within which\
quantities like density and pressure vary by several orders of magnitude
from the bottom to the top.
\item The largest relevant scales ($10^6$ km) differ from the 
smallest scales of fibril flux tubes ($10^2$ km) by 4 orders. 
\end{itemize}
Additionally, a kinematic model has one tremendously important advantage.
Within the last few years, helioseismology has provided considerable
information about the velocity fields in the Sun's interior and it
has now been possible to put these
velocity fields measured by helioseismology directly into the
kinematic dynamo models, thereby revolutionizing the field. In
a DNS, on the other hand, the velocity fields have to be calculated
from the basic equations of fluid mechanics and, until one gets
the velocity fields correctly, there is no hope of getting the magnetic
fields correctly.  Of late, several research groups are again working
on DNS of the solar dynamo problem (Ghizaru, Charbonneau \&
Smolarkiewicz 2010; Brown et al.\ 2010).  It is possible that a major
breakthrough is around the corner. However, until such a breakthrough
takes places, the DNS models are still of rather exploratory nature.
If one seeks theoretical explanations of various detailed aspects
of the solar cycle, so far the kinematic mean-field model is the
only possible approach. Also, we hope that results of kinematic
mean-field models will provide useful guidance in the development
of DNS models.

\section{Flux transport dynamo model}

The central idea of solar dynamo theory is that the toroidal and the poloidal
components of the magnetic field sustain each other through a feedback
loop.  It is easy to see how the poloidal field may give rise to the
toroidal field. The differential rotation of the Sun is expected to
stretch the poloidal field lines in the toroidal direction.  Since
the differential rotation is strongest in the tachocline at the base
of the convection zone, we expect the toroidal field to be primarily
produced there.  This toroidal field then has to rise through the
convection zone due to magnetic buoyancy to produce sunspots.  Numerical
simulations of this buoyant rise suggested that the magnetic field at
the bottom of the convection zone has to be as strong as $10^5$ G
(Choudhuri \& Gilman 1987; Choudhuri 1989; D'Silva \& Choudhuri 1993;
Fan, Fisher \& DeLuca 1993).

The generation of the poloidal field from the toroidal field is a more
complicated problem.  Historically there have been two influential schools
of thought.  The original idea of Parker (1955) and Steenbeck, Krause
\& R\"adler (1966) is that turbulence in the presence of rotation has
a preferred helicity and this helical turbulence can twist the toroidal
field to produce the poloidal field.  This mechanism is often called
the $\alpha$-effect. The alternative idea due to Babcock (1961) and
Leighton (1964) is based on the fact the bipolar sunspot pairs on the
solar surface appear with a tilt, produced by the action of the Coriolis
force on the rising flux tubes (D'Silva \& Choudhuri 1993). 
After the decay of tilted bipolar sunspots, their
magnetic fields diffuse around to give rise to a poloidal field. A tilted
bipolar sunspot pair can thus be viewed as a conduit for converting the
toroidal field to the poloidal field.  It forms from the toroidal field
and we get the poloidal field after its decay. 

Many of the early solar dynamo calculations were based on the $\alpha$-effect.
However, the $\alpha$-effect can twist the toroidal field only if it has
a value not much stronger than the value that would produce equipartition
between the energies of turbulence and magnetic field.  The equipartition
value of the magnetic field at the bottom of the convection zone is
expected to be not larger than $10^4$ G.  When flux rise simulations indicated
that the toroidal field may be as strong as $10^5$ G, it appeared that the
$\alpha$-effect will not be able to twist such a strong field. Hence the
Babcock--Leighton mechanism is invoked in many recent models. Since we see 
this mechanism operating on the solar surface, we believe this to be the
dominant mechanism under normal circumstances. But, during the grand minima
when sunspots are absent, the Babcock--Leighton mechanism may not take place.
We probably need something like an $\alpha$-effect to pull the Sun out of
a grand minimum.  How the Sun comes out of a grand minimum is very poorly
understood at the present time. The discussion in this review will be
restricted to normal situations when the Babcock--Leighton mechanism is
presumably the primary mechanism for generating the poloidal field.

\begin{figure}
\center
\includegraphics[width=7cm]{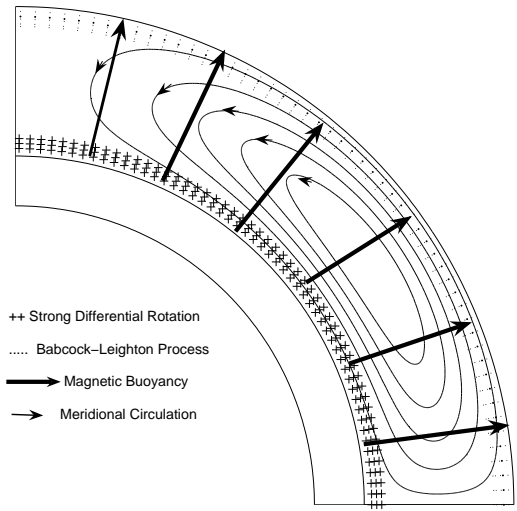}
  \caption{A cartoon explaining how the flux transport dynamo works.}
\end{figure}

\def\mc{meridional circulation}

We can consider a dynamo in which the toroidal field is produced by differential
rotation in the tachocline and the poloidal field is produced near the
surface by the Babcock--Leighton mechanism.  Choudhuri, Sch\"ussler \&
Dikpati (1995), who studied this kind of dynamo, pointed out one possible
difficulty. According to the Parker-Yoshimura sign rule (Parker 1955; Yoshimura
1975; Choudhuri 1998, \S16.6), a dynamo based solely on these effects will
have a poleward propagation, contrary to the observations.  However, Choudhuri,
Sch\"ussler \& Dikpati (1995) showed that a suitable meridional circulation
can reverse the propagation direction and can give rise to equator-propagating
butterfly diagrams.  Dynamo models in which the meridional circulation plays
an important role are called flux transport dynamo.  Such a dynamo was
considered in an early paper 
by Wang, Sheeley \& Nash (1991).  First two-dimensional models
of the flux transport dynamo were constructed by Choudhuri, Sch\"ussler
\& Dikpati (1995) and Durney (1995).

Fig.~1 shows a schematic cartoon of the flux transport dynamo.  The
meridional circulation which is equatorward at the bottom of the convection
zone advects the toroidal field produced there.  On the other hand,
the poleward meridional circulation near the surface advects the poloidal
field poleward, which is seen observationally.  Fig.~2 shows the
butterfly diagram of sunspots in the same plot along with the
time-latitude plot of the radial field at the surface. The theoretical
plot obtained by Chatterjee, Nandy \& Choudhuri (2004) shown on
the right has to be compared with the observational plot shown on the
left. Most of the calculations of the flux transport dynamo model
have been done with single-celled meridional circulation which
reach out to the bottom of the convection zone, as shown in
Fig.~1. In fact, it is found that certain aspects of observational
data can be explained best if the \mc\ is assumed to penetrate
slightly below the bottom of the convection zone (Nandy \& Choudhuri
2002; Chakraborty, Choudhuri \& Chatterjee 2009).
There are some recent indications that the meridional circulation
may be more complicated than this and the return flow may be at
a shallower depth (Hathaway 2012; Junwei Zhao, private communication). 
It remains to be seen whether these results get confirmed by other 
independent studies. Flux transport dynamo models with more complicated
\mc\ are yet to be studied properly.

\begin{figure}
\begin{minipage}[c]{0.55\textwidth}
\vspace{0.1cm}
\includegraphics[height=6.5cm,width=9.5cm]{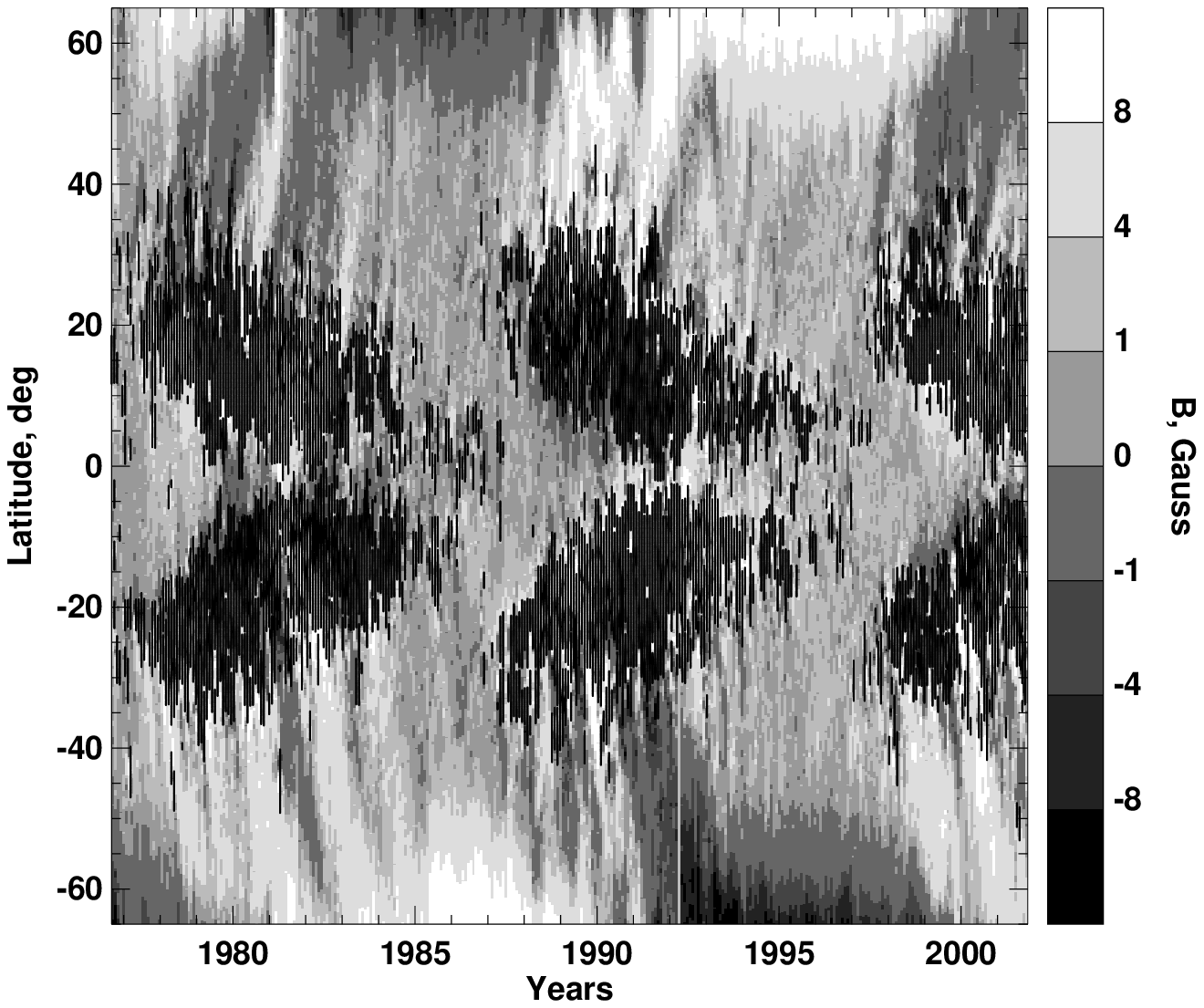}
\end{minipage}%
\hspace{0.3cm}
\begin{minipage}[c]{0.6\textwidth}
\includegraphics[height=6cm,width=6.0cm]{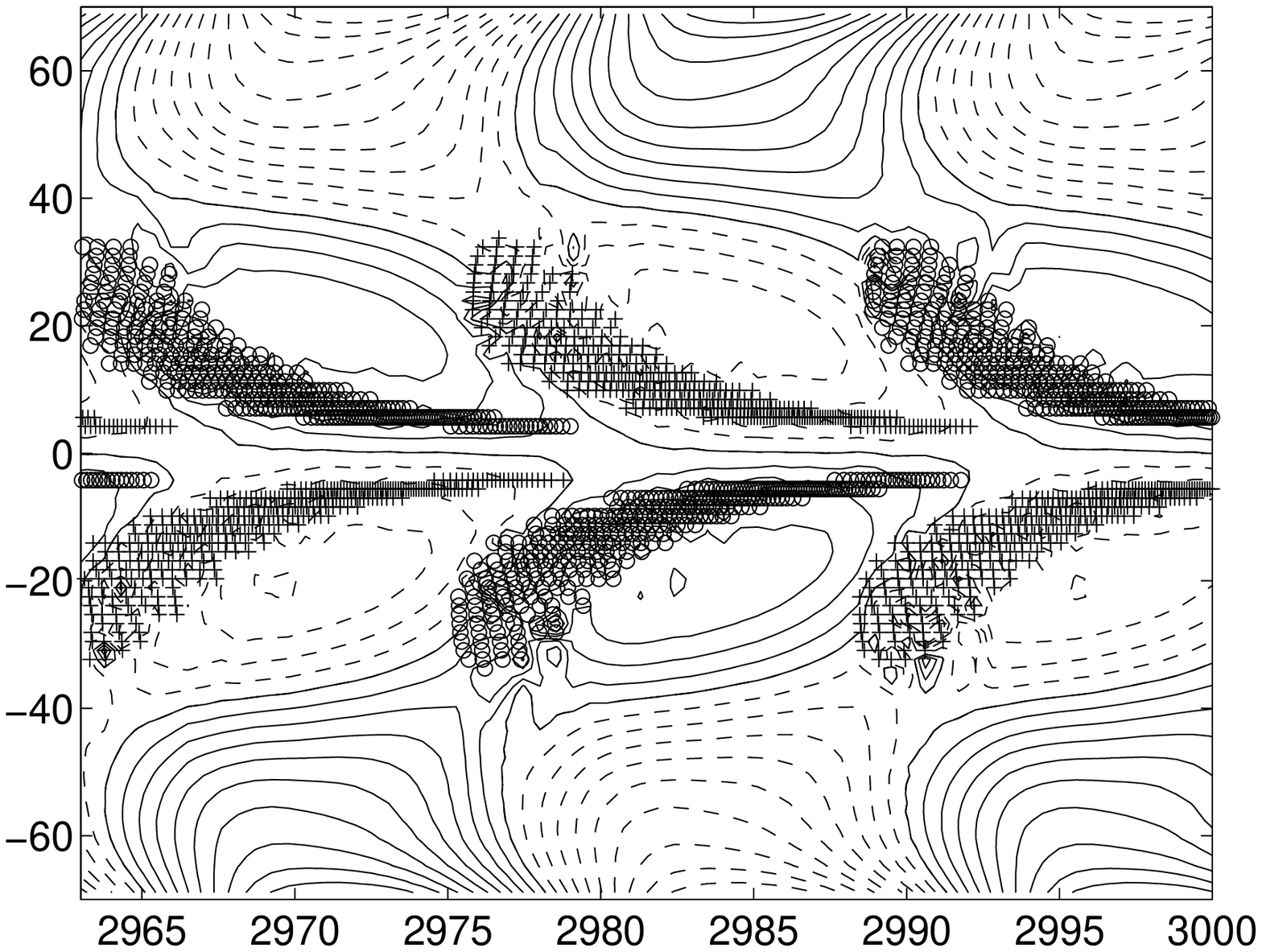}
\end{minipage}
\caption{Butterfly diagram of sunspots superposed on the time-latitude
  plot of $B_r$. The observational plot is shown on the left.  
  The comparable theoretical
  plot obtained by the dynamo model of Chatterjee, Nandy \& Choudhuri 
(2004) is on the right.}
\end{figure}

While the flux transport dynamo model, as sketched in Fig.~1, may not
yet be a universally accepted model of the solar cycle, more and more
scientists are finding it an attractive model and considerable work has
been done on this model in the last few years
by different groups around the world. Given the
limited scope of this review, it is not possible to cover all the
aspects of the flux transport dynamo model currently under investigation.
The next two sections provide a rather personal and subjective sampling
of topics which are of special interest to this author and on which our
group had been working. Notwithstanding the disclaimer that this is 
not a comprehensive
review of the flux transport dynamo, hopefully readers will get an idea
of some of the crucial issues. Some issues connected with the flux
transport dynamo not discussed here are discussed by Jiang (2013).
 
\section{Two classes of models and the problem of modelling 
irregularities of the solar cycle}

In the flux transport dynamo model, the toroidal field is produced at
the bottom of the convection zone, whereas the poloidal field is produced
at the top.  We need some transport mechanisms between these two source
regions in order for the dynamo to work. As shown in Fig.~1, the toroidal field is transported
by magnetic buoyancy from the bottom to the top where the Babcock--Leighton
mechanism acts on it to produce the poloidal field.  We also need a transport
mechanism for bringing the poloidal field from the top to the bottom where
differential rotation can act on it to produce the toroidal field.  Two
classes of models have been worked out in the last few years based on two
different dominant transport mechanisms for this.  If the turbulent diffusivity of
the convection zone is assumed to be sufficiently high to make the diffusion
time across the convection zone of order 5 years, then the poloidal field
reaches the bottom from the top by diffusion. Such high-diffusivity models
have been constructed in Bangalore by Choudhuri and his successive students
(Nandy, Chatterjee, Jiang, Karak).  On the other hand, Dikpati and her
co-workers in Boulder (Charbonneau, Gilman, de Toma)
have developed a low-diffusivity model in which the
diffusion time scale across the convection zone is of order hundreds of years
and the poloidal field is advected from the top to the bottom by the
meridional circulation shown in Fig.~1. The differences between these models
have been systematically studied by Jiang, Chatterjee \& Choudhuri
(2007) and Yeates, Nandy \& Mckay (2008).  Both these models are
capable of giving rise to oscillatory solutions resembling solar
cycles.  However, when we try to study the irregularities of the
cycles, the two models give completely different results.  We need
to introduce fluctuations to cause irregularities in the cycles.
In the high-diffusivity model, the fluctuations spread all over the
convection zone in about 5 years.  On the other hand, in the 
low-diffusivity model, the fluctuations essentially remain frozen during
the cycle period.  Thus the behaviours of the two models are totally
different on introducing fluctuations.  

It may be mentioned that both these models were used a few years ago
to predict the strength of the cycle 24 before it had begun.  Dikpati
\& Gilman (2006) used their low-diffusivity model to predict that the
cycle~24 will have the peak sunspot number in the range 157--181, 
making it one of the strongest recorded cycles. On
the other hand, Choudhuri, Chatterjee \& Jiang (2007) used their 
high-diffusivity model to predict a peak sunspot number of around 70--80,
implying that it will be a weak cycle.
While the jury is not completely out yet, the early indications suggest
that the prediction from the high-diffusivity model will be much closer
to the truth.  This lends more credibility to the high-diffusivity
model. We list below several other arguments in favour of the 
high-diffusivity model.
\begin{itemize}
\item The diffusivity assumed in the high-diffusivity model is consistent
with the simple mixing length estimate $\frac{1}{3} v l$ (Parker 1979, p.\ 629; Miesch
et al.\ 2012).
\item The high diffusivity helps in establishing dipolar parity which
the solar magnetic fields seem to have (Chatterjee, Nandy \& Choudhuri 2004; 
Hotta \& Yokoyama 2010).
\item The high diffusivity keeps the hemispheric asymmetry in solar activity 
small as observed 
(Chatterjee \& Choudhuri 2006; Goel \& Choudhuri 2009).
\item The surface flux transport models also require a similar diffusivity 
near the surface (Wang, Nash \& Sheeley 1989).
\item The high-diffusivity model reproduces the observed correlation between the 
polar field at the minimum and the strength of the next cycle 
(Jiang, Chatterjee \& Choudhuri 2007).
\end{itemize}
The last point is crucial in the prediction of forthcoming cycles. In the
work on predicting cycle~24, Choudhuri, Chatterjee \&
Jiang (2007) had fed the information in the dynamo model
that the polar field during the preceding minimum was weak and it is an
inevitable consequence of the high-diffusivity model that the next cycle
has to be weak as a result of the last point (Jiang, Chatterjee \& Choudhuri 
2007). Apart from the arguments
listed above, we shall see in the next section that the high-diffusivity model
is able to explain certain other aspects of observational data 
(such as the Waldmeier effect) on introducing
fluctuations in \mc.

Let us mention here that, apart from meridional circulation and diffusion,
there is another possible mechanism for transporting the poloidal field from
the top of the convection zone to the bottom: turbulent pumping. Only
recently effects of this on the flux transport dynamo have started being studied (Karak \&
Nandy 2012; Jiang et al.\ 2012). Since the downward transport time scale due to 
pumping is comparable to the diffusion time scale
in the high-diffusivity model, the results on inclusion
of pumping are qualitatively similar to the results of the high-diffusivity
model, even when the diffusivity is assumed to be low.

In recent years, a major thrust in the research on solar dynamo has been to
understand how irregularities arise in solar cycles.  Both the high-diffusivity
and low-diffusivity models have been used to study the irregularities.  The
irregularities in the solar cycle are believed to be caused by the following
three mechanisms.
\begin{enumerate}
\item Nonlinear effects giving rise to possible chaotic behaviour.
\item Fluctuations in the Babcock--Leighton mechanism for poloidal field
generation.
\item Fluctuations in the \mc.
\end{enumerate}
It is beyond the scope of this review to discuss all these three mechanisms
in detail.
We shall concentrate here only on the last mechanism which is being studied
systematically within the last couple of years, making just a few
brief remarks on the first two mechanisms. An earlier review by
Choudhuri (2012) discusses all the three mechanisms.  For a more complete
discussion on the fluctuations in the Babcock--Leighton mechanism, see a still
earlier review (Choudhuri 2011). Presumably the nonlinearities are responsible
for such things as the Gnevyshev--Ohl effect (Charbonneau, Beaubien \& St-Jean 2007).
Since the Babcock--Leighton mechanism depends on the tilt angles of bipolar
sunspot pairs and these tilt angles show a scatter around the mean given by
Joy's law, presumably due to the effect of turbulence on the rising flux
tubes (Longcope \& Choudhuri 2002), we expect some inherent randomness in
the Babcock--Leighton mechanism.   When predicting the strength of cycle~24,
Choudhuri, Chatterjee \& Jiang (2007) assumed this to be the main cause of
irregularities in the solar cycle. At that time, the important role played
by fluctuations in the \mc\ was not yet appreciated.  Presumably the prediction method
developed by Choudhuri, Chatterjee \& Jiang (2007) will be reliable only if
there is no sudden variation in the \mc\ between the time when the prediction
was made and the time when the cycle eventually reaches its peak.

\begin{figure}
\centering
\includegraphics[width=1.00\textwidth]{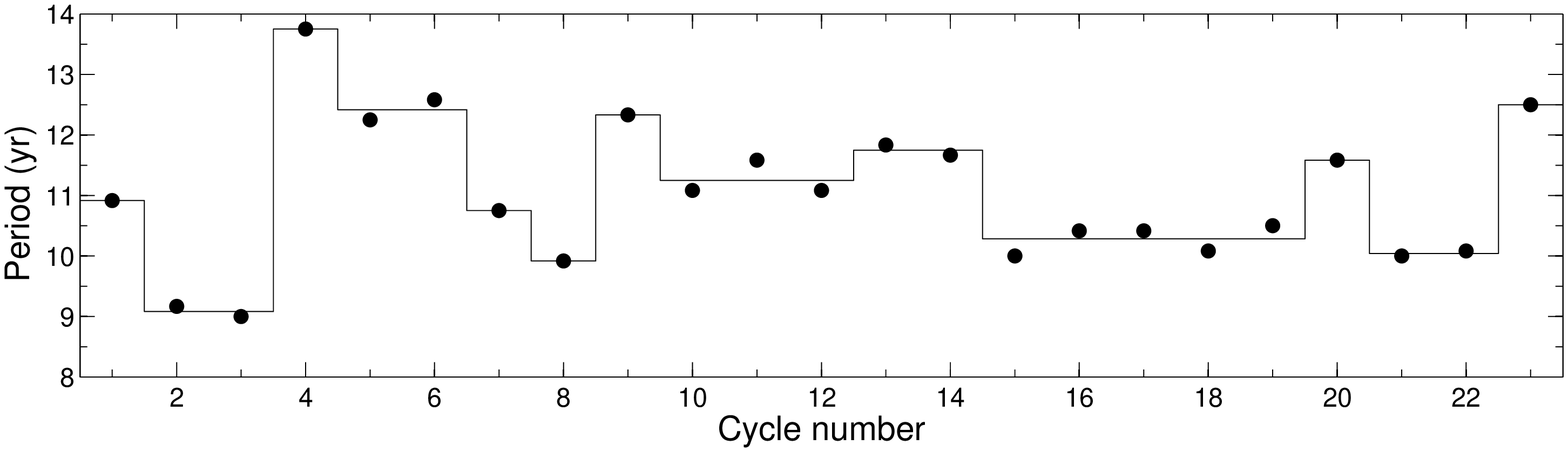}
\caption{The points show the periods of last 23 solar cycles against 
cycle number. The solid line is indicative of the trend in variations
of period.}
\label{period}
\end{figure}

\section{Fluctuations in meridional circulation}

It is well known that the period of the flux transport dynamo varies roughly as
the inverse of the \mc\ speed.  A slower \mc\ would make the cycle longer.
It is, therefore, obvious that fluctuations
in \mc\ would have some effect on the flux transport dynamo. Recently it has been found
that the \mc\ has a periodic variation with the solar cycle, becoming
weaker at the time of sunspot maximum (Hathaway \& Rightmire 2010; Basu \& Antia 2010).
Presumably the Lorentz force of the dynamo-generated magnetic field slows down
the \mc\ at the time of the sunspot maximum.  Karak \& Choudhuri (2012) found
that this quenching of \mc\ by the Lorentz force does not produce irregularities
in the cycle, provided the diffusivity is high as we believe.  We disagree with
the model of Nandy, Mu\~noz-Jaramillo \& Martens (2011) which assumes that the 
\mc\ changes randomly at each sunspot maximum.  Our point of view is that the
periodic variation of \mc\ due to the Lorentz force cannot be responsible for
randomness in solar cycles and we need to consider other kinds of fluctuations
in \mc.

We have reliable observational data on the variations of \mc\ only for a little
more than a decade.  To draw any conclusion about the variations of \mc\ at
earlier times, we have to rely on indirect arguments. If we assume the cycle
period to go inversely as \mc, then we can use periods of different past solar
cycles to infer how \mc\ has varied with time in the last few centuries. 
Fig.~3 plots periods of several past cycles.  We note that several successive
cycles had short periods, indicating that the \mc\ was probably fast at
that time.  On the other hand, some successive cycles had longer periods,
implying a slower \mc. On the
basis of such considerations, it appears that the \mc\ had random fluctuations
in the last few centuries with correlation time of the order of 30--40 years (Karak \& Choudhuri 2011).
We now come to question what effect these random fluctuations of \mc\ may have
on the dynamo.  Based on the analysis of Yeates, Nandy \& Mckay (2008), we can
easily see that high-diffusivity and low-diffusivity dynamos will be affected very
differently.  Suppose the \mc\ has suddenly fallen to a low value. This will
increase the period of the dynamo and lead to two opposing effects.  On the one
hand, the differential rotation will have more time to generate the toroidal field and
will try to make the cycles stronger.  On the other hand, diffusion will also have
more time to act on the magnetic fields and will try to make the cycles weaker.
Which of these two competing effects wins over will depend on the value of diffusivity.
If the diffusivity is high, then the action of diffusivity is more important and
the cycles become weaker when the \mc\ is slower.  The opposite happens if the
diffusivity is low.

\begin{figure}
\centering
\includegraphics[width=1.0\textwidth]{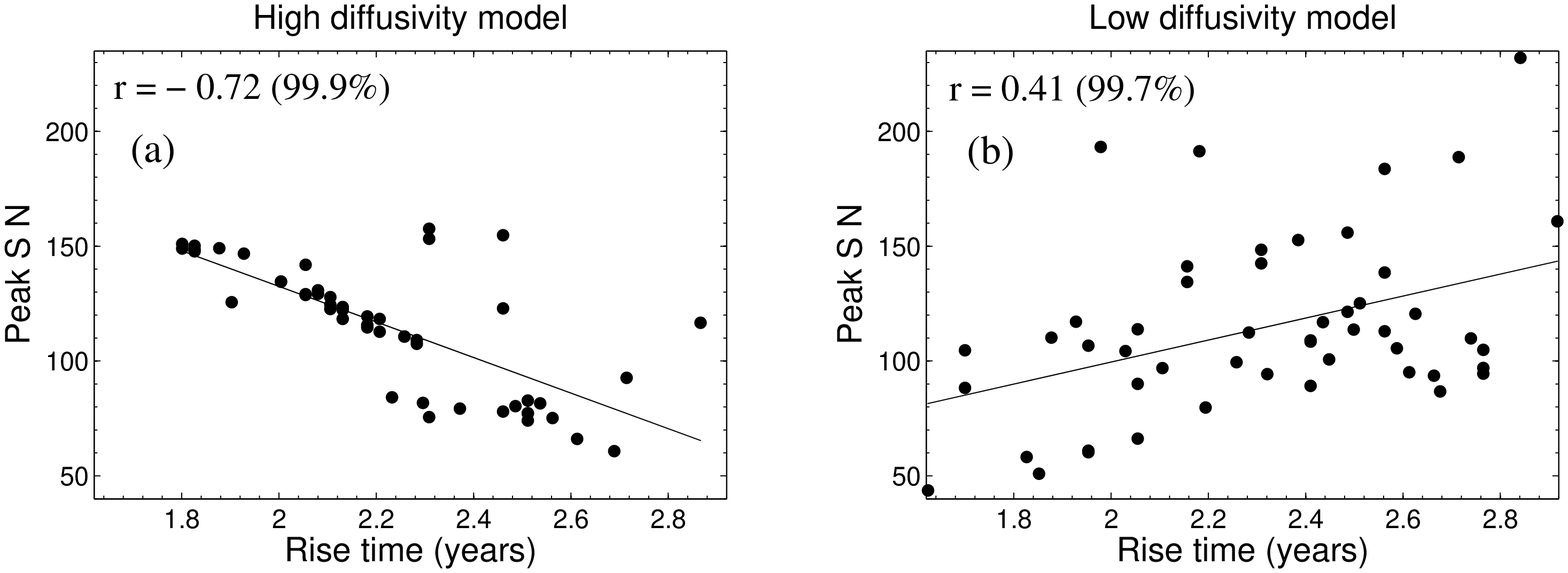}
\vspace{-5cm}
\caption{Theoretical plots of correlation between the
peak sunspot number and the rise time (in years), obtained by introducing fluctuations 
in the meridional circulation. The two figures correspond to the
high-diffusivity and the low-diffusivity models. Taken from Karak
\& Choudhuri (2011).}
\end{figure}

We now address the question if we can compare our theoretical conclusions with
any observational data.  It has been known for a long time that stronger cycles
take shorter time to rise (Waldmeier 1935). In other words, there is an 
anti-correlation between the rise times and strengths of cycles. This is
known as the Waldmeier effect.  The
discussion in the last paragraph shows how this effect can be explained
with the high-diffusivity dynamo model, which was first done by Karak \&
Choudhuri (2011). When the \mc\ slows down, the period becomes longer and simultaneously
the cycle becomes weaker in the case of the high-diffusivity dynamo. This
means that we would have an anti-correlation between the periods and the
cycle strengths. Such an anti-correlation naturally leads also to an 
anti-correlation between rise times and cycle strengths --- the Waldmeier
effect.  A little reflection will convince you that the opposite --- a
direct correlation between rise times and cycle strengths --- will happen
in the low-diffusivity model.  Although Charbonneau \& Dikpati (2000)
considered fluctuations in \mc\ having much shorter correlation times, still
their Fig.~4(c) shows a direct correlation between the cycle duration and
cycle amplitude. Charbonneau, Beaubien \& St-Jean (2007) 
made the following comment about the Waldmeier effect:
``This has remained notoriously difficult to reproduce within extant solar 
cycle models.'' These authors found it so difficult to reproduce the
Waldmeier effect because they were using the low-diffusivity model with
which it seems impossible to explain the Waldmeier effect.  On using
the high-diffusivity model, the Waldmeier effect almost falls on your
lap, as found by Karak \& Choudhuri (2011). Fig.~4 shows correlation
plots between rise time and peak sunspot number for both high-diffusivity
and low-diffusivity models on introducing fluctuations in \mc. Only the
high-diffusivity model produces the Waldmeier effect.

Some of the variations in the cycle strength during the last few
cycles seem to be due to fluctuations in \mc.  Karak (2010) carried
out an interesting simulation in which he varied the \mc\ to match
only the periods of past cycles and found that even the strengths
of many cycles got approximately modelled in this process.  Fig.~5
taken from Karak (2010) shows the actual sunspot number by the solid
line, whereas the dashed line is the sunspot number obtained from
the theoretical model by varying the \mc\ to match the periods.
On carrying out the same exercise with the low-diffusivity model,
Karak (2010) found that the cycle strengths are not matched at all.
Even in the case of the high-diffusivity model, we do not expect the cycles
to be matched completely on varying the \mc\ because fluctuations
in the Babcock--Leighton mechanism also contribute to the 
irregularities of the cycles.

One of the most intriguing aspects of the solar cycle is
grand minima when several cycles go missing. Although
the only well-documented grand minimum to 
occur after telescopic observations began is the Maunder 
minimum during 1640-1715, indirect proxies provide
evidence for 27 such grand minima in the 
last 11,000 years (Usoskin, Solanki \& Kovaltsov 2007). Since
the slowing of the \mc\ in the high-diffusivity model makes the
cycles weaker, one crucial question is whether the Sun can be pushed
into a grand minimum if the \mc\ becomes sufficiently weak due
to its intrinsic fluctuations.  Karak (2010) showed that this is
indeed possible.  However, a grand minimum can also occur when
the poloidal field falls to very low values due to fluctuations
in the Babcock--Leighton mechanism (Choudhuri \& Karak 2009).
To study the occurrence of grand minima and to calculate its
probability, it is necessary to consider fluctuations in the
\mc\ and in the Babcock--Leighton mechanism simultaneously.
This has now been done by Choudhuri \& Karak (2012). Since the
theory of grand minima is discussed in another paper in this
Proceedings volume (Karak \& Choudhuri 2013), we do not get into
the details of this subject here and refer the reader to this
other paper.

\begin{figure}
\centering
\includegraphics[width=1.0\textwidth]{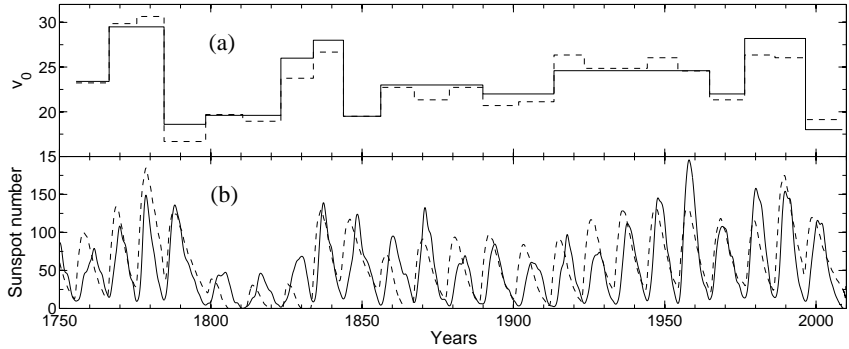}
\vspace{-4cm}
\caption{The top panel shows the variation of the \mc\ amplitude
$v_0$ (in m s$^{-1}$) obtained from the periods of the past
cycles by assuming the period to go as $\propto v_0^{-0.696}$
(dashed line), whereas the solid line indicates $v_0$ actually
used in the theoretical simulations to get the best fit for
periods of past cycles. The bottom panel shows the theoretically
calculated sunspot number (dashed line) along with the observational
sunspot number (solid line). Taken from Karak (2010). }
\end{figure}

\section{Conclusion}

Although several full simulations of the solar dynamo are currently under way,
the kinematic mean-field models so far remain the primary theoretical tools for
explaining different aspects of the solar cycle in detail. The most promising
kinematic mean-field model at the present time is the flux transport dynamo
model, in which the toroidal field is generated by the differential rotation
at the bottom of the convection zone, the poloidal field is generated near the
solar surface by the Babcock--Leighton
mechanism and the \mc\ plays a crucial role by ensuring that magnetic fields
are transported in the appropriate directions.  In the class of models in which
the diffusivity is assumed low, the \mc\ turns out to be also the primary mechanism
for transporting the poloidal field from the surface to the bottom of the
convection zone.  On the other hand, in the other class of models with high
diffusivity, it is the diffusivity which causes the transport of the poloidal
field from the top to the bottom.  We have argued that the high-diffusivity
flux transport dynamo model is consistent with various aspects of the observational
data and is most probably the appropriate model for the solar cycle.

Modelling irregularities of the solar cycle is an active area of research
right now.  There are three primary mechanisms which may give rise to the
irregularities. We have made some comments about two of these mechanisms,
nonlinear effects and fluctuations in the Babcock--Leighton mechanism, but
have refrained from discussing them in detail due to the limitation of space
and also because they have been discussed elsewhere. The third mechanism,
fluctuations in the \mc, which is being studied systematically only during the
last couple of years, is discussed more fully.  Introducing these fluctuations
in the high-diffusivity model, we are able to explain such things as the
Waldmeier effect.  We believe that the fluctuations in the Babcock--Leighton
mechanism and in the \mc\ jointly give rise to the various irregularities
of the solar cycle, including the grand minima.

\section*{Acknowledgments}

My participation in IAU Symposium 294 was made possible by a JC Bose Fellowship 
awarded by Department of Science and Technology, Government of India.

\end{document}